\documentclass[aps,prx,twocolumn,superscriptaddress,nofootinbib,notitlepage,longbibliography]{revtex4-1}
\usepackage{bm}
\usepackage{hhline}
\usepackage{amsmath}
\usepackage{amssymb}
\usepackage{mathdots}
\usepackage{tabularx}
\usepackage{graphicx}
\usepackage{siunitx}
\usepackage{hyperref}
\usepackage{bbding}
\usepackage{tikz}
\usepackage{bbding}
\usepackage{braket}
\usepackage{bm}

\begin{document}
	\title{Orbital Longitudinal Magnetoelectricity in Quasi-2D Parity-Violating Antiferromagnets: Interplay of Berry Phase and Stacking Order}
	\author{Jin-Xin Hu}\thanks{jhuphy@ust.hk}
    \affiliation{Division of Physics and Applied Physics, School of Physical and Mathematical Sciences, Nanyang Technological University, Singapore 637371}  
    
        \affiliation{Department of Physics, Hong Kong University of Science and Technology, Clear Water Bay, Hong Kong, China} 


	\begin{abstract}

In this work, we propose an intrinsic mechanism for longitudinal magnetoelectric (ME) coupling in quasi-2D antiferromagnets that lack spatial inversion symmetry. We demonstrate that in such parity-violating antiferromagnets, the stacking order, when coupled with Berry curvature and orbital magnetic moment, generates a stacking Berry curvature dipole (SBCD) and a stacking orbital magnetic moment dipole (SOMD). The SBCD and SOMD act as the fundamental ingredients of the longitudinal ME response. As concrete examples, we apply our framework to typical parity-violating antiferromagnets with stacking magnetic orders. Furthermore, the SBCD is also identified as the origin of electric-field-induced Hall effects. Our results reveal the microscopic orbital origin of the ME effect in antiferromagnetic systems with vanishing net Berry curvature and orbital magnetization, governed by the interplay between Berry phase effects and layer stacking orders.

	\end{abstract}
	\maketitle

\section{Introduction}
The magnetoelectric (ME) effect describes the generation of magnetization by an electric field and of electric polarization by a magnetic field~\cite{catalan2006magnetocapacitance,gupta2022review,mori2002magnetoelectric,katsura2007dynamical}. This cross-coupling offers a promising paradigm for next-generation electronic devices, enabling efficient control of magnetic and electric orders, which is a central objective for low-energy-consumption spintronics~\cite{fusil2014magnetoelectric,bea2008spintronics,trassin2015low}. Historically, the ME effect has been extensively studied in three-dimensional (3D) single-phase multiferroics such as Cr$_2$O$_3$~\cite{iyama2013magnetoelectric} and BiFeO$_3$~\cite{wang2003epitaxial,kuo2016single}, where it typically arises from relativistic spin-orbit coupling and complex spin-lattice interactions~\cite{zhu2017modulation,lee2010strong}. In recent years, interests have grown in the topological origin of the ME12 effect, including the gyrotropic ME effect~\cite{zhong2016gyrotropic,he2020magnetoelectric} and topological ME responses~\cite{essin2009magnetoelectric,wang2015quantized}. 

The recent discovery of intrinsic magnetism in van der Waals (vdW) materials, such as CrI$_3$ and MnBi$_2$Te$_4$, has opened a new frontier in exploring low-dimensional magnetism~\cite{sivadas2018stacking,jiang2019stacking,sun2019giant,lei2021magnetoelectric,deng2020quantum,li2019intrinsic,otrokov2019unique,li2024dissipationless,gao2024antiferromagnetic,gao2021layer,gao2023quantum,wang2023quantum}. A key feature of these quasi-2D systems is their unique layer stacking order, which typically breaks spatial inversion symmetry ($\mathcal{P}$). When combined with magnetic order that breaks time-reversal symmetry ($\mathcal{T}$), this stacking magnetic order establishes the fundamental prerequisite for linear ME effect~\cite{liu2020magnetoelectric,qiu2025observation}. The modern theory of the ME effect focuses on axion coupling in 3D systems. For instance, the ME coefficient ishalf-quantized to $e^2/2h$ in axion insulators~\cite{lei2024capacitive,mei2024electrically}. In parallel, the role of Berry phase effects in electronic transport, such as the anomalous Hall effect~\cite{nagaosa2010anomalous,xiao2010berry}, its nonlinear counterpart~\cite{sodemann2015quantum} and orbital magnetization~\cite{shi2007quantum,ceresoli2006orbital}, has been well studied. Up to date, the role of the Berry phase in linear ME phenomena, particularly in quasi-2D magnetic systems, remains largely unexplored.

\begin{figure}
		\centering
		\includegraphics[width=1.0\linewidth]{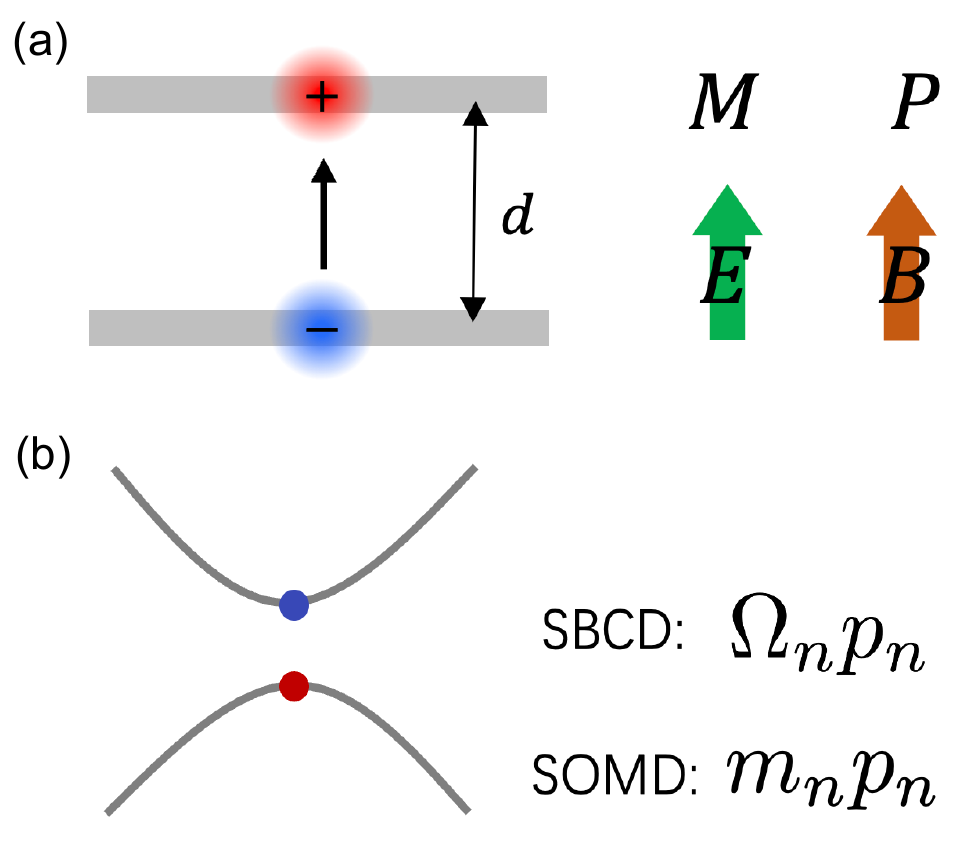}
		\caption{(a) Schematic illustration of the longitudinal ME effect in a bilayer system. The interlayer distance is $d$. Under the applied magnetic ($B$) field and electric ($E$) field, the electric polarization and magnetization can be induced. (b) The stacking Berry curvature dipole (SBCD) and stacking orbital magnetic moment dipole (SOMD) are centralized near the band edge, giving rise to the orbital ME effect.}
		\label{fig:fig1}
\end{figure}

In this work, we elucidate the orbital origin of magnetoelectricity in non-centrosymmetric quasi-2D antiferromagnetic materials with vanishing total Berry curvature and orbital magnetization. Focusing on systems with combined symmetries such as $\mathcal{PT}$ symmetry, we demonstrate that the stacking Berry curvature dipole (SBCD) and the stacking orbital magnetic dipole (SOMD) are two universal quantities for intrinsic magnetoelectricity in such systems (see Fig.~\ref{fig:fig1}). We posit that the layer stacking order creates a synthetic dimension in which a dipole moment of Berry curvature (orbital magnetic moment) can develop; this dipole is directly modulated by the stacking magnetic order. This mechanism provides a purely electronic, intrinsic origin for the ME effect, distinct from other extrinsic mechanisms. We provide a comprehensive picture of longitudinal ME effects in vdW magnets, and the general formalism can be readily incorporated in DFT calculations.

This work is structured as follows. First, we present a general microscopic theory for longitudinal ME coupling in quasi-2D systems. We emphasize that, unlike in multilayer graphene, the SBCD and SOMD govern the longitudinal ME coupling in antiferromagnetic materials with vanishing net Berry curvature and orbital magnetization. Second, we apply our general theory to an antiferromagnetic toy model with $C_{4z}\mathcal{PT}$ symmetry, which can describe a monolayer of Ca(CoN)$_2$~\cite{zhang2024predictable}. Third, we study the ME effect in topological antiferromagnetic multilayers using the prototypical material MnBi$_2$Te$_4$, where finite interlayer coupling causes the ME coefficient in the insulating regime to deviate from the half-quantized value. Furthermore, we propose that in certain $\mathcal{PT}$-symmetric antiferromagnets such as tetragonal CuMnAs, the longitudinal ME effect can be controlled by tuning the spatial orientation of the N\'{e}el order. Finally, we summarize our findings and discuss the distinct microscopic origins of the ME effect across these van der Waals (vdW) magnets. This work establishes a comprehensive picture of longitudinal ME effects in vdW heterostructures, and the general formalism can be readily incorporated into first-principles calculations.

\section{General theory of longitudinal ME coupling}
The first-order linear longitudinal ME coupling is described microscopically by $\delta M = \chi_{\text{me}} E$ and $\delta P = \chi_{\text{em}} B$, where $\delta M$ is the induced magnetization, $\delta P$ is the induced polarization, and $E$ and $B$ are the applied out-of-plane electric and magnetic fields, respectively. The ME coefficients are denoted $\chi_{\text{me}}$ and $\chi_{\text{em}}$, and a Maxwell relation ensures $\chi_{\text{em}} = \chi_{\text{me}}$. For simplicity we take $\Delta=ed E$ to denote the vertical electric field.

For quasi-2D layered systems, the linear ME coefficient is generally given by~\cite{hu2026orbital}
\begin{eqnarray}
   \chi_{\text{me}} &  = e \sum_{n\bm{k}} \Big\{-\alpha_{n\bm{k}} f_{n\bm{k}} +  m_{n\bm{k}} p_{n\bm{k}} f'_{n\bm{k}}  \nonumber \\ 
     &+ \frac{e}{\hbar} \big[\frac{1}{\beta}\delta_\Delta \Omega_{n\bm{k}} \phi (\varepsilon_{n\bm{k}}) - \Omega_{n\bm{k}} f_{n\bm{k}} p_{n\bm{k}}\big] \Big\},
   \label{eq:generalchi_me}
\end{eqnarray}
where $n,\bm{k}$ are the band index and wavevector, $\phi(\varepsilon)=\mathrm{ln}[1+e^{-\beta(\varepsilon-\mu)}]$ and $f_{n\bm{k}} = -\partial_\epsilon \phi(\epsilon)/\beta = \{1+ {\rm exp}{[\beta (\varepsilon_{n\bm{k}} -\mu})]\}^{-1}$ is the Fermi-Dirac function. The Berry curvature is $\Omega_{n\bm{k}}=i\sum_{m\neq n}\bm{r}_{nm}\times \bm{r}_{mn}$ and the orbital magnetic moment is  $m_{n\bm{k}} = (e/2)\sum_{m\neq n}\bm{v}_{nm}\times \bm{r}_{mn}$ with $\bm{v}_{nm}$ and $\bm{r}_{nm}$ the interband velocity matrix elements and interband Berry connection respectively. The first term in Eq.~\eqref{eq:generalchi_me} is so called the orbital ME moment $\alpha_{n\bm{k}}\approx -\delta_\Delta m_{n\bm{k}}$ that governs ME effect in multilayer graphene~\cite{hu2026orbital}. 

The layer dipole operator $\hat{p}$ is generally given by $\hat{p}=\sum_{\ell,\alpha} (\ell/d)  |\phi_{\ell,\alpha} \rangle \langle \phi_{\ell,\alpha}|$, where $ |\phi_{\ell,\alpha} \rangle$ denotes the $\alpha$ orbital on the layer at height $\ell$ measured from the center of the stack. For example, for a bilayer system $\ell = (d/2) \{-1,1\}$ and $\hat{p}=\sigma_z/2$. In the band summation form, $\delta_\Delta m_{n\bm{k}}$ and $\delta_\Delta \Omega_{n\bm{k}}$ can be obtained as
\begin{equation}
\label{eq:par_mnk}
\begin{split}
  \delta_\Delta m_{n\bm{k}}=-\sum_{m\neq n}\frac{e\hbar}{\varepsilon_{nm}^2}&\mathrm{Im}[\varepsilon_{nm}\delta_\Delta\bm{v}_{nm}\times \bm{v}_{mn}+\\
  & v_{nm}^x v_{mn}^y (p_m-p_n)],
\end{split}    
\end{equation}
and
\begin{equation}
\label{eq:par_onk}
\begin{split}
  \delta_\Delta \Omega_{n\bm{k}}=-\sum_{m\neq n}\frac{2\hbar^2}{\varepsilon_{nm}^3}&\mathrm{Im}[\varepsilon_{nm}\delta_\Delta\bm{v}_{nm}\times \bm{v}_{mn}+\\
  & 2v_{nm}^x v_{mn}^y (p_m-p_n)],
\end{split}    
\end{equation}
where $ \delta_\Delta \bm{v}_{nm}= \sum_{q\neq n}p_{nq}\bm{v}_{qm}/\varepsilon_{nq}+\sum_{l\neq m}p_{lm}\bm{v}_{nl}/\varepsilon_{ml}$. For antiferromagnetic systems with vanishing net Berry curvature and orbital magnetization, $\delta_\Delta m_{n\bm{k}}$ as well as $\delta_\Delta \Omega_{n\bm{k}}$ terms are zero since they are the full derivative forms. Therefore, $\chi_{\text{me}}$ has the form 
\begin{equation}
\label{eq:eq_ptme_t}
   \chi_{\text{me}}  = e \sum_{n} \int \frac{d^2\bm{k}}{(2\pi)^2}  (-\frac{e}{\hbar} \mathcal{B}_{n\bm{k}} f_{n\bm{k}}+\mathcal{M}_{n\bm{k}} f'_{n\bm{k}}),
\end{equation}
where we have defined the SBCD $\mathcal{B}_{n\bm{k}}=\Omega_{n\bm{k}}p_{n\bm{k}}$ and SOMD $\mathcal{M}_{n\bm{k}}=m_{n\bm{k}}p_{n\bm{k}}$. Eq.~\eqref{eq:eq_ptme_t} is the main result of this work. It is clear that the first term is Fermi-sea effect governed by $\mathcal{B}_{n\bm{k}}$, while the second term is Fermi-surface effect governed by $\mathcal{M}_{n\bm{k}}$. It is important to note that both $\mathcal{B}_{n\bm{k}}$ and $\mathcal{M}_{n\bm{k}}$ survives $\mathcal{PT}$ symmetry. 

\begin{figure}
		\centering
		\includegraphics[width=1.0\linewidth]{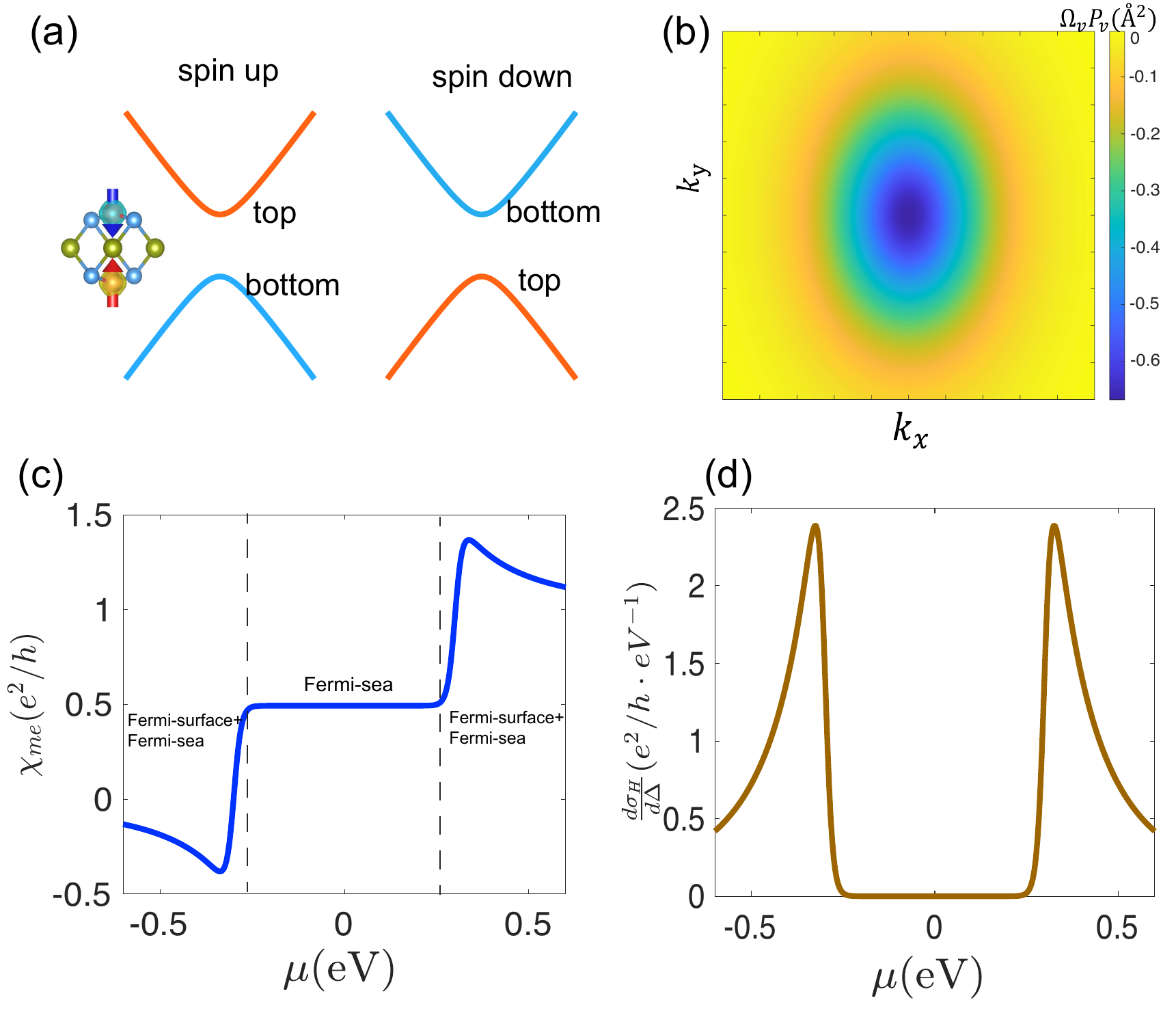}
		\caption{(a) Schematic of the low-energy band structure with $C_{4z}\mathcal{PT}$ symmetry, illustrating the spin-layer locking. The $X$ and $Y$ valleys correspond to spin-up and spin-down states, respectively. The side panel shows magnetic moments localized mainly on the top and bottom Co atoms, aligned in opposite directions. (b) Momentum-space distribution of the SBCD $\mathcal{B}_{v\bm{k}}$ with hot spot near $\Gamma$ pocket. (c) Calculated ME coefficient $\chi_{\text{me}}$ as a function of Fermi energy $\mu$. (d) Response of the electric Hall conductivity, $d\sigma_{H}/d\Delta$. Parameters for the model: $(v_1, v_2) = (0.6, 0.4)$ $\mathrm{eV}\cdot \AA$ and $U = 0.3$ eV.}
		\label{fig:fig2}
\end{figure}
\begin{figure*}
		\centering
		\includegraphics[width=1.0\linewidth]{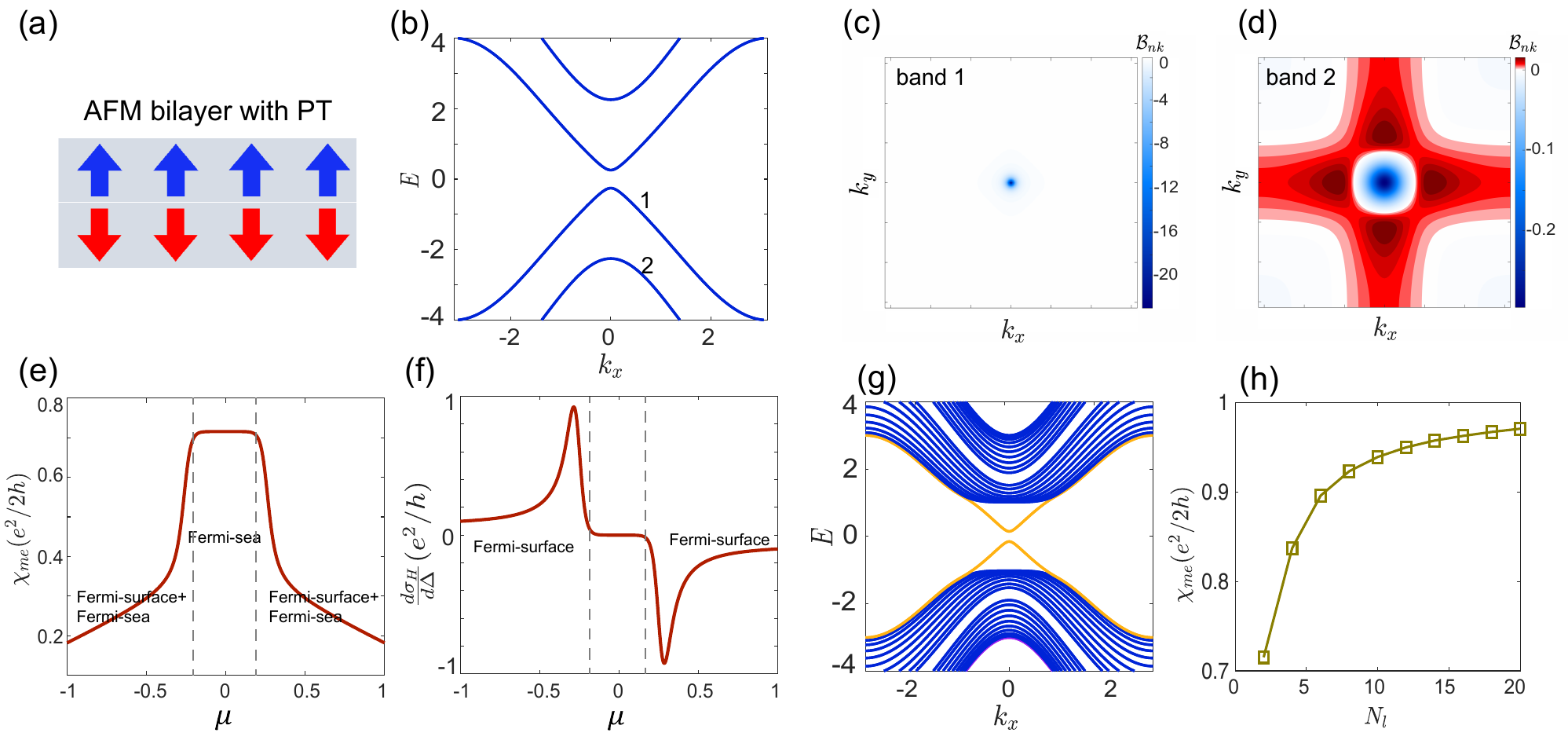}
		\caption{(a) Schematic plot of the AFM bilayers wth $\mathcal{PT}$ symmetry and interlayer AFM order. (b) The band structure along $k_x$ and $k_y=0$ for $N_l=2$. (c) and (d) The $k$-space SBCD $B_{n\bm{k}}$ for the band 1 and 2 in (b). (e) The ME coeifficient $\chi_{me}$ as a function if $\mu$ in the low-energy regime. Both Fermi-surface and Fermi-sea terms are shown. (f) The electric Hall effect $d\sigma_H/d\Delta$ as a function if $\mu$. (g) The band structure of multilayer AFM with $N_l=20$. The gapped surface state is labeled by orange lines. (h) The $\chi_{me}$ approaches $e^2/2h$ as the (even) number of layer goes up. Parameters: $A=1.4,B=1,M_0=1,m_z=0.3$. The temperature is set to be $k_B T=0.02$.}
		\label{fig:fig3}
\end{figure*}

In antiferromagnets with $\mathcal{PT}$ symmetry, the anomalous Hall effect vanishes because the net Berry curvature is zero. The vertical electric field breaks $\mathcal{PT}$ producing net anomalous Hall conductivity. We can derive the electric Hall effect
\begin{equation}
\label{eq:eq_elec_Hall}
\frac{d\sigma_{H}}{d\Delta}=\frac{e^2}{\hbar}\sum_n \int \frac{d^2\bm{k}}{(2\pi)^2}(\mathcal{B}_{n\bm{k}}f'_{n\bm{k}}+\delta_\Delta\Omega_{n\bm{k}}f_{n\bm{k}})
\end{equation}
It describes how the anomalous Hall effect can be induced by the electric fields~\cite{gao2021layer,cui2025electric,han2025layer}. The second term in Eq,~\eqref{eq:eq_elec_Hall} vanishes for the antiferromagnets with zero Berry curvature as discussed previously.

\section{Example 1: $C_4 \mathcal{PT}$-symmetric antiferromagnet}
To illustrate the ME effect induced by the SBCD and SOMD, we employ the low-energy effective model of monolayer Ca(CoN)$_2$~\cite{zhang2024predictable}. As shown in Fig.~\ref{fig:fig2}(a), the electronic states in different valleys exhibit opposite layer polarization. The electronic states with opposite spin polarization (residing in different valleys) also possess opposite layer polarization, leading to a valley-mediated spin-layer locking. For monolayer Ca(CoN)$2$, the magnetic point group of the $X$ and $Y$ valleys is $m'm'2$. Within the basis $(\varphi_{t},\varphi_{b})$, the effective Hamiltonian at a given valley can be expressed as
\begin{eqnarray}
H_X &=& U \tau_z+ v_1 k_x \tau_x+v_2 k_y \tau_y \\
H_Y &=& -U \tau_z+ v_1 k_y \tau_x-v_2 k_x \tau_y
\end{eqnarray}
Here the $X$ and $Y$ valleys are related by $C_{4z}\mathcal{PT}$ symmetry. We can analytically derive the SBCD as well as SOMD for each valley separately. For $X$ valley, the SBCD for the conduction/valence band is given by
\begin{equation}
\mathcal{B}_{c/v}=-\frac{v_1v_2 U^2}{2(U^2+v_1^2 k_x^2+v_2^2 k_y^2)^2}
\end{equation}
and the SOMD for the conduction/valence band is given by
\begin{equation}
\mathcal{M}_{c/v}=\mp\frac{e}{2\hbar}\frac{v_1v_2 U}{(U^2+v_1^2 k_x^2+v_2^2 k_y^2)^{3/2}}
\end{equation}
The quantities $\mathcal{B}$ and $\mathcal{M}$ have the same signs for the $Y$ valley due to $C_{4z}\mathcal{PT}$ symmetry. In Fig.~\ref{fig:fig2}(b), we plot the momentum-space distribution of $\mathcal{B}_{v\bm{k}}$ for the $X$ valley. The integral of $\mathcal{B}_{v\bm{k}}$ yields the ME coefficient $\chi_{\text{me}}=e^2/2h$. Note that this value is for a continuum model where the momentum integral extends to infinity; in a real tight-binding model with a periodic Brillouin zone, $\chi_{\text{me}}$ will be smaller.

As displayed in Fig.~\ref{fig:fig2}(c), we evaluate $\chi_{\text{me}}$ as a function of Fermi energy $\mu$. In the insulating regime, $\chi_{\text{me}}$ is solely contributed by the Fermi-sea term $\mathcal{B}_{\bm{k}}$, while in the metallic regime, the Fermi-surface term $\mathcal{M}_{\bm{k}}$ also arises. Both of them are intrinsic and the inherent properties of a given model Hamiltonian. The sign of the Fermi-surface term is opposite for the conduction and valence bands because the orbital magnetic moment has the same sign in both, leading to $\mathcal{M}_{c\bm{k}}=-\mathcal{M}_{v\bm{k}}$. In contrast, the electric Hall effect is purely a Fermi-surface effect governed by the SBCD $\mathcal{B}_{\bm{k}}$, as shown in Fig.~\ref{fig:fig3}(d). In this case, the sign of $d\sigma_H/d\Delta$ is the same for the conduction and valence bands, since $\mathcal{B}_{c\bm{k}}=\mathcal{B}_{v\bm{k}}$. It is worth noting that the spin magnetic moment can also give rise to the ME coupling as a Fermi-surface effect, while in this work we focus on the orbital origin.

\section{Example 2: topological antiferromagnet multilayers.}
To further explore the longitudinal ME effect and electric Hall effect, we take a second example by examining a bilayer magnetic insulator, as shown in Fig.~\ref{fig:fig3}(a). The corresponding tight-binding model Hamiltonian for 3D topological insulator can be expressed as~\cite{fu2007topological,jiang2012quantum}
\begin{equation}\label{Ham}
  H_0=\sum_{i=1,2,3,4,5}d_{i}(\bm{k})\Gamma_i
\end{equation}
where $\bm{d}(\bm{k})=(-A\sin k_x,-A\sin k_y, 0, M_0-6B+2B(\cos k_x +\cos k_y+\cos k_z),-A\sin k_z)$ and $\bm{\Gamma}=(\sigma_x s_x,\sigma_x s_y,\sigma_y s_0,\sigma_z s_0,\sigma_x s_z)$. Here $\sigma$ and $s$ denote the orbital and spin degree of freedom. In a thin film with finite thickness, the Hamiltonian for each layer is 
\begin{equation}
\begin{split}
h_0&=-A(\sin k_x \Gamma_1+\sin k_y \Gamma_2)\\
&+[M_0-6B+2B(\cos k_x+\cos k_y)]\Gamma_4
\end{split}
\end{equation}
and the interlayer coupling term $T=-iA\Gamma_5/2+B\Gamma_4$. Therefore, the model Hamiltonian for the antiferromagnet bilayer is given by
\begin{equation}\label{eq:bdg}
H_{tb}=\left(\begin{array}{cc}
h_0+m_z \sigma_0 s_z& T \\
T^\dagger & h_0-m_z \sigma_0 s_z
\end{array}\right).
\end{equation}
where $m_z$ denotes the layer-dependent exchange field characterizing the interlayer antiferromagnetic order. This model can describe the essential physics of bilayer MnBi$_2$Te$_4$~\cite{zhang2019topological}. Throughout this work we take the parameters $A=1.4,B=1,M_0=1,m_z=0.3$ to illustrate the core physics.

We first study the bilayer case (\(N_l=2\)), as illustrated in Fig.~\ref{fig:fig3}(a). The band structure is shown in Fig.~\ref{fig:fig3}(b), with valence bands labeled 1 and 2. The momentum-space distribution of the SBCD for bands 1 and 2 is plotted in Fig.~\ref{fig:fig3}(c) and (d), respectively. This reveals that band 1 provides the dominant contribution to the orbital ME coupling, while the effect from band 2 is minor. This distinction can be understood from the significantly larger layer polarization of band 1. Using Eq.~\eqref{eq:eq_ptme_t}, we evaluate \(\chi_{\text{me}}\) as a function of the Fermi level \(\mu\) in Fig.~\ref{fig:fig3}(e). The SBCD contributes to a Fermi-sea ME coupling with a non-quantized value of \(\chi_{\text{me}}  \approx 0.72\). This deviation from quantization occurs because finite interlayer coupling results in a layer polarization \(\langle p \rangle \neq 1/2\) for each layer. Notably, \(\chi_{\text{me}}\) has the same sign for the conduction and valence bands because \(\mathcal{M}_{c\bm{k}}=\mathcal{M}_{v\bm{k}}\) in this system. We further calculate the electric Hall effect (or layer Hall effect) in Fig.~\ref{fig:fig3}(f). As expected for a Fermi-surface effect, \(d\sigma_{H}/d\Delta\) stems from the SBCD and vanishes in the insulating regime. Finally, we consider systems beyond the bilayer. Figure~\ref{fig:fig3}(g) shows the band structure for \(N_l=20\), where the low-energy gapped surface states induced by \(m_z\) are highlighted in orange. In Fig.~\ref{fig:fig3}(h), we plot the evolution of the ME coefficient \(\chi_{\text{me}}\) with increasing layer number \(N_l\). We observe that \(\chi_{\text{me}}\) clearly approaches the quantized value of \(e^2/2h\) in the large-thickness limit.

\begin{figure}
		\centering
		\includegraphics[width=1.0\linewidth]{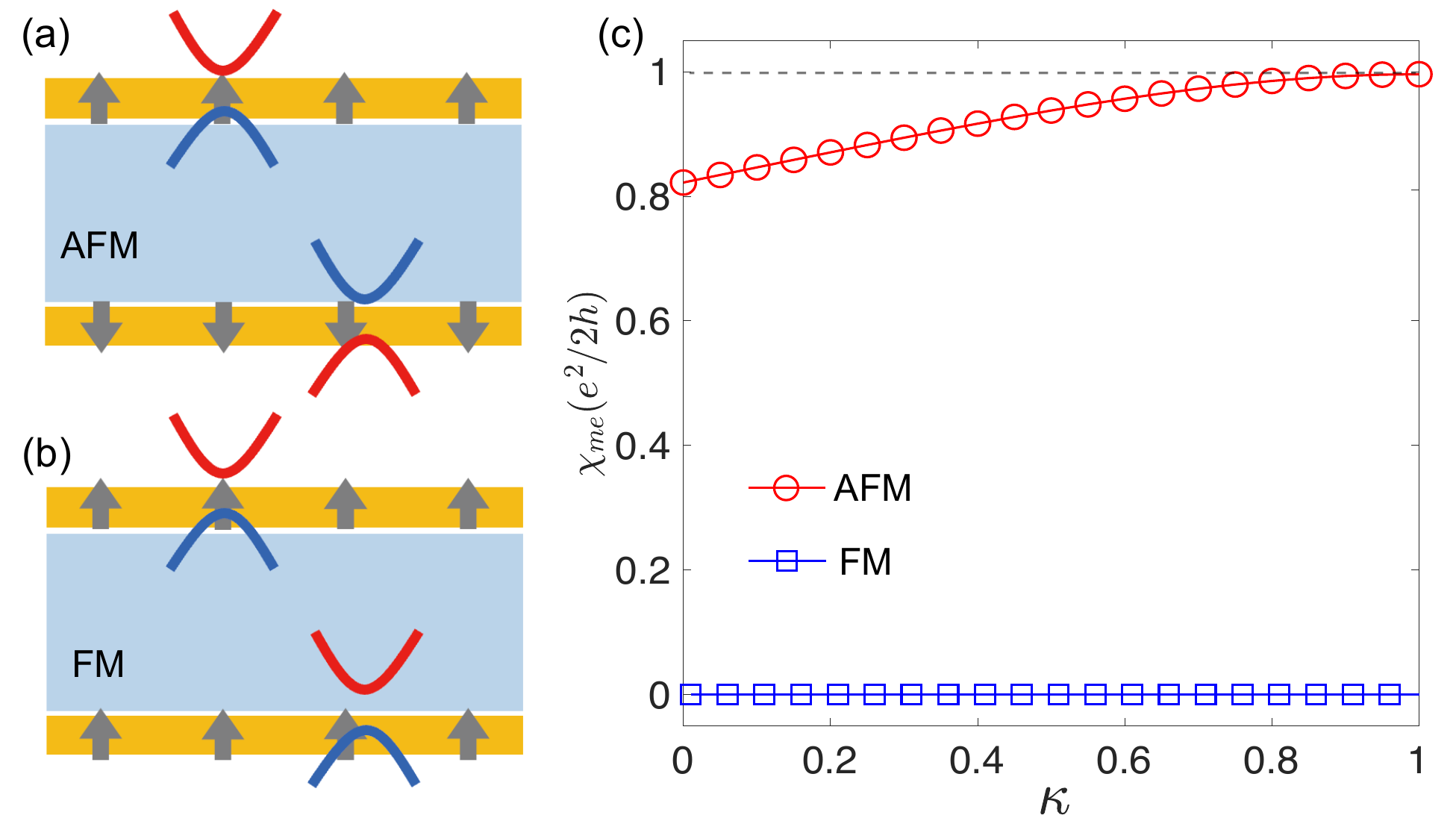}
		\caption{(a) Schematic for an axion insulator with gapped surface state. The Dirac mass of top and bottom layers are opposite. (b) The FM bilayer with the quantum anomalous Hall state with the same Dirac mass for two layers. (c) $\chi_{me}$ as a function of $\kappa$. When $\kappa=1$, the interlayer coupling vanishes.}
		\label{fig:fig4}
\end{figure}

Multilayer magnetic topological insulators (MTI) have the gapped surface state localized on the top and bottom surfaces. Therefore, we can adopt an effective ``bilayer" Hamiltonian, yielding
\begin{equation}
H_{eff}=\hbar v_0(k_ys_x-k_xs_y)\tau_z+(1-\kappa)m(\bm{k})\tau_x+h_M
\end{equation}
Here we take $\hbar v_0=2 $ $\mathrm{eV}\cdot \AA$. The interlayer coupling is given by $m(\bm{k}) = m_0 - m_1(k_x^2 + k_y^2)$ with parameters $m_0 = -10\,\text{meV}$ and $m_1 = 20\,\text{eV}\cdot \text{\AA}^2$. Here, $s$ and $\tau$ are the Pauli matrices acting on the spin and layer subspaces, respectively. For the antiferromagnetic (AFM) bilayer, the magnetization term is $h_M = m_z s_z \tau_z$, while for the ferromagnetic (FM) bilayer it is $h_M = m_z s_z \tau_0$. This leads to opposite Dirac masses in the AFM case and identical Dirac masses in the FM case, as illustrated in Fig.~\ref{fig:fig4}(a) and (b). We note that the AFM bilayer models even-layer MTIs, whereas the FM bilayer models odd-layer MTIs~\cite{mei2024electrically}. The parameter $\kappa$ controls the interlayer coupling strength, where $\kappa=1$ corresponds to effectively decoupled layers.

Focusing on the Fermi-sea term from $\mathcal{B}_{n\bm{k}}$, we evaluate $\chi_{\text{me}}$ for both AFM and FM cases. As shown in Fig.~\ref{fig:fig4}(c), $\chi_{\text{me}}$ for the AFM bilayer saturates to the half-quantized value $e^2/2h$ as $\kappa$ approaches 1, where the two layers are effectively decoupled. This quantization arises because the Berry curvature integral for a massive Dirac cone on the top/bottom layer is $\pm 1/2$, and the corresponding layer polarization is $\langle p \rangle = \pm 1/2$. Substituting these values into Eq.~\eqref{eq:eq_ptme_t} directly yields $\chi_{\text{me}} = e^2/2h$. In contrast, $\chi_{\text{me}}$ vanishes for the FM bilayer for all values of $\kappa$, which occurs because the SBCD contributions from the top and bottom layers are equal in magnitude but opposite in sign, leading to perfect cancellation.

\section{Example 3: PT-symmetric AFM with tunable Neel order}
\begin{figure}
		\centering
		\includegraphics[width=1.0\linewidth]{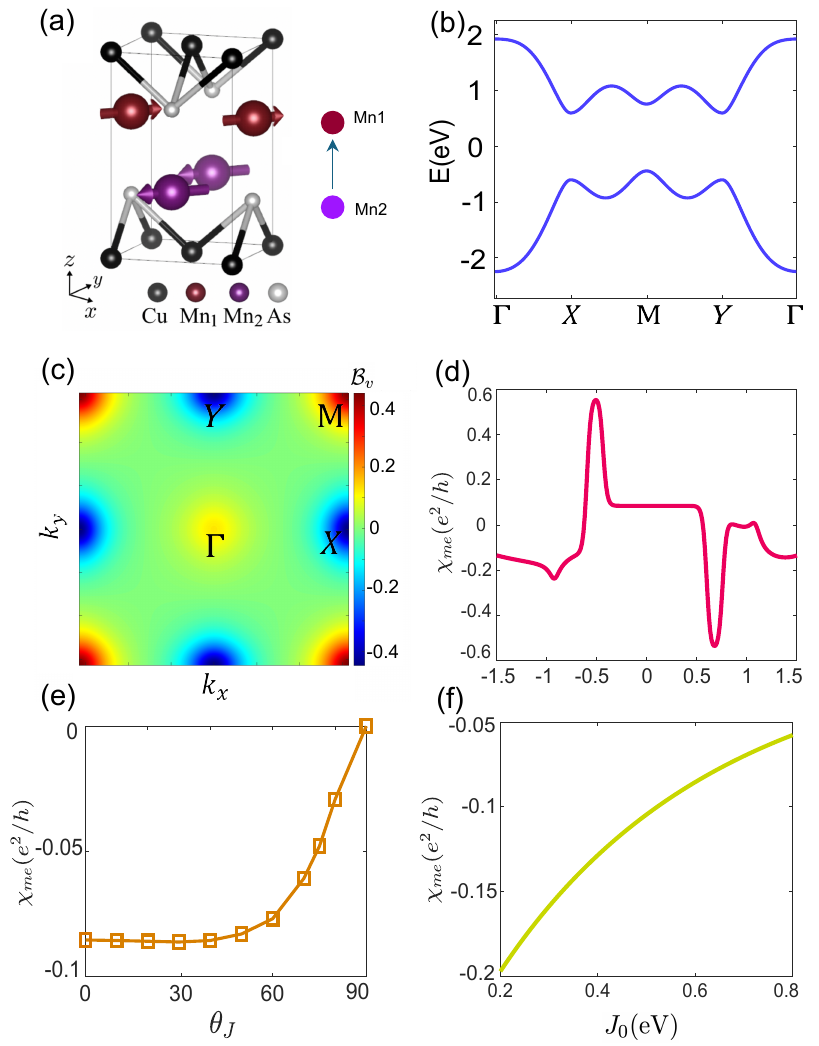}
		\caption{(a) The crystal lattice structure of quasi-2D tetragonal CuMnAs with an antiferromagnetic order. The two Mn (Mn1 and Mn2) atoms can formulate a vertical dipole moment. (b) The band structure along high symmetry line at $\theta_J=0$. (c) Momentum-space distribution of the SBCD $\mathcal{B}_{v\bm{k}}$ (d) Calculated ME coefficient $\chi_{\text{me}}$ as a function of Fermi energy $\mu$. (e) and (f) The angle ($\theta_J=0$) and magnitude ($J_0$) dependence of $\chi_{\text{me}}$.}
		\label{fig:fig5}
\end{figure}

In this section, we consider a quasi-2D antiferromagnetic material, tetragonal CuMnAs. In its antiferromagnetic phase, the magnetic moments on two Mn sites connected by inversion symmetry are oriented opposite to each other. This configuration breaks spatial inversion ($\mathcal{P}$) and time-reversal ($\mathcal{T}$) symmetries individually but preserves their product, $\mathcal{PT}$ symmetry~\cite{godinho2018electrically,wadley2016electrical,vsmejkal2017electric}, as illustrated in Fig.~\ref{fig:fig5}(a). Crucially, the two Mn atoms can form an out-of-plane electric dipole moment along the $z$-direction. As we show later, this dipole moment is essential for generating a longitudinal magnetoelectric response associated with Berry curvature. The single-layer, quasi-2D structure of tetragonal CuMnAs can be modeled by the following tight-binding Hamiltonian~\cite{vsmejkal2017electric}
\begin{equation}
\label{eq:Jungwirth}
\begin{split}
H(\bm{k}) &= -2t \sigma_{x} \cos \frac{k_{x} a}{2} \cos \frac{k_{y} a}{2} - t' \left( \cos k_{x} a + \cos k_{y} a \right) \\
&+\lambda  \sigma_{z}  \left( s_{y} \sin k_{x} a - s_{x} \sin k_{y} a \right) + \sigma_{z} J_{0} \vec{s} \cdot \vec{n},
\end{split}
\end{equation}
where $t$ is the nearest-neighbor hopping, $t'$ is the next-nearest-neighbor hopping, $\lambda$ is the next-nearest-neighbor spin-orbit coupling, $\vec{n}$ is the N\'{e}el vector, and the Pauli matrices $\bm{\sigma}$ and $\bm{s}$ act on the sublattice and spin degrees of freedom, respectively. In this representation, the inversion operator is $P = \sigma_x s_0$, and the time-reversal operator is $T = -i \sigma_0 s_y K$, where $K$ denotes complex conjugation. The final term in Eq.~\eqref{eq:Jungwirth} breaks both $\mathcal{P}$ and $\mathcal{T}$ individually but preserves the combined $\mathcal{PT}$ symmetry. The parameters are set as $\lambda = 0.8t$, $J_{n} = 0.6t$, $t' = 0.08t$, and $|t| = 1\,\text{eV}$. The band structure depends sensitively on the orientation of the N\'{e}el vector, $\vec{n}=(\sin\theta_J\cos\varphi_J,\sin\theta_J\sin\varphi_J,\cos\theta_J)$~\cite{vsmejkal2017electric}. The vertical electric dipole moment is modeled as $\hat{p} = (d_z/2) \sigma_z$, where $d_z$ is the vertical distance between the two Mn atoms.

Within the band structure calculated at $\theta_J=0$ in Fig.~\ref{fig:fig5}(b), we plot the momentum-space distribution of the SBCD in Fig.~\ref{fig:fig5}(c). The hotspots of $\mathcal{B}_{v\bm{k}}$ are located around the $X$, $M$, and $Y$ pockets. We then calculate $\chi{\text{me}}$ as a function of the Fermi level $\mu$ in Fig.~\ref{fig:fig5}(d), which exhibits similar Fermi-sea and Fermi-surface behaviors as discussed in previous examples. Notably, the Fermi-sea contribution yields a value of $\chi_{\text{me}} = -0.083,e^2/h$, which varies strongly with the AFM order parameter $\theta_J$, as shown in Fig.~\ref{fig:fig5}(e). This tunability arises from the strong dependence of the SBCD term $\mathcal{B}_{\bm{k}}$ on $\theta_J$, demonstrating that stacking geometric quantities are closely related to the AFM order in $\mathcal{PT}$-symmetric antiferromagnets. Furthermore, $\chi_{\text{me}}$ depends strongly on the exchange parameter $J_0$, as displayed in Fig.~\ref{fig:fig5}(f). We note that a large derivative $\delta \chi_{\text{me}} / \delta J_0$ may indicate the presence of dynamical axion quasiparticles~\cite{qiu2025observation}.

Finally, we would like to connect our theory to the topological ME effect within Chern-Simons theory. The topological ME response in an antiferromagnetic topological insulator is described by the topological $\theta$-term
\begin{equation}
    S_{\theta} = \frac{\theta}{2\pi} \frac{e^2}{h} \int d^3x\, dt\, \bm{E} \cdot \bm{B},
\end{equation}
where $\bm{E}$ and $\bm{B}$ are the electromagnetic fields inside the insulator, and $\theta$ is a dimensionless pseudoscalar parameter defined modulo $2\pi$. Physically, $\theta$ has an explicit microscopic expression as a momentum-space Chern-Simons form~\cite{essin2009magnetoelectric}. This parameter describes three fundamental physical phenomena: (i) The electric polarization induced by a small magnetic field. (ii) The orbital magnetization induced by a small electric field. (iii) A quantized, dissipationless surface Hall conductivity $\sigma_H = e^2/2h$.

In contrast, for quasi-2D layered systems, we find that an analogous $\theta_{\text{2D}}$ is determined by the integrated SBCD
\begin{equation}
    \theta_{\text{2D}} = \sum_{n \in \text{occ}} \int d^2\bm{k}\, \mathcal{B}_{n\bm{k}},
\end{equation}
which is a relation that holds generally for quasi-2D parity-violating antiferromagnetic insulators when the electromagnetic fields are applied along vertical direction. While the surface Hall effect is not well-defined in atomically thin systems, we show that $\mathcal{B}_{n\bm{k}}$ nonetheless gives rise to a measurable electric Hall effect, as derived in Eq.~\eqref{eq:eq_elec_Hall}. 

Strikingly, a recent experiment has observed dynamical axion quasiparticles in quasi-2D even-layer MnBi$_2$Te$_4$~\cite{qiu2025observation}. In this system, coherent oscillations of the axion angle $\theta$ are driven by out-of-phase antiferromagnetic magnons under an in-plane magnetic field. Beyond a static $\theta$, the dynamical $\delta\theta(t)$ is induced by magnetic fluctuations. A larger $\delta\theta/\delta L_z$ (where $L_z$ is the AFM order parameter) produces a stronger signature of these dynamical axion quasiparticles. As our analysis indicates, this condition can be achieved in typical quasi-2D AFMs such as CuMnAs.

\begin{figure}
		\centering
		\includegraphics[width=1.0\linewidth]{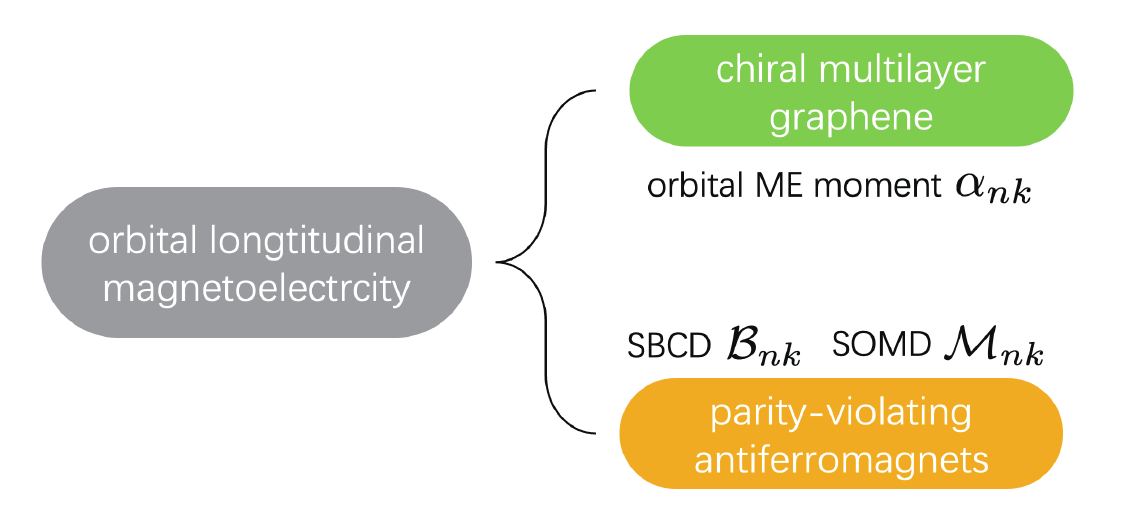}
		\caption{Summary of the origins of orbital magnetoelectricity in two representative systems: in chiral-stacked multilayer graphene, the effect is governed by the orbital ME moment $\alpha_{n\bm{k}}$, while in parity-violating antiferromagnets, it is dominated by the SBCD and SOMD, which is  the primary focus of this work.}
		\label{fig:fig6}
\end{figure}

\emph{Conclusion and discussion.}---In summary, in this work we have studied that the interplay between Berry phase effects and stacking order in quasi-2D layered materials gives rise to a distinct orbital ME effect, differing fundamentally from known mechanisms in 3D systems. We have identified the SBCD and SOMD as the key quantities governing this ME response in parity-violating antiferromagnets. This theoretical framework was demonstrated using two illustrative models.

As shown in Fig.~\ref{fig:fig6} and derived from the general formula of Eq.~\eqref{eq:generalchi_me}, we have clarified the different microscopic origins of the orbital ME effect. In chiral multilayer graphene systems, the orbital ME moment $\alpha_{n\bm{k}}$ determines the longitudinal ME coupling, as studied in Ref.~\cite{hu2026orbital} and supported by relevant experiments~\cite{han2023orbital,deng2025superconductivity}. In contrast, the SBCD and SOMD govern the longitudinal ME coupling in antiferromagnets such as Ca(CoN)$_2$ and MnBi$_2$Te$_4$. This work establishes a unified picture of the orbital ME effect in layered van der Waals materials. The general formalism developed here can be readily incorporated into first-principles calculations, providing a pathway for the quantitative prediction and design of ME phenomena in low-dimensional quantum materials.

\emph{Acknowledgement.}---We are grateful to Justin C. W. Song, Yugui Yao, and X .C. Xie for fruitful discussions. J.X.H. was supported by the postdoctoral fellowship of Nanyang Technological University and Hong Kong University of Science and Technology.

	\appendix
	\renewcommand{\theequation}{A-\arabic{equation}}
	\renewcommand\thefigure{A-\arabic{figure}}
	\setcounter{equation}{0}
	\setcounter{figure}{0}
	
\section{Microscopic theory of orbial longitudinal magnetoelectric effect}\label{AppendixA}
Here we present the general theory of longtitudinal ME effect, which can be applied to quasi-2D model Hamiltonian. We can start from the field-dependent free energy $F(E,B)$, which reads
\begin{equation}
\label{eq:free_energy}
F(E,B)=-\frac{1}{\beta}\sum_{n\bm{k}}\big\{1+\frac{e}{\hbar}B\tilde{\Omega}_{n\bm{k}}\big\}\phi\big[\varepsilon_{n\bm{k}} (E,B)\big],
\end{equation}
where $n,\bm{k}$ are the band index and wavevector, $\phi(\varepsilon)=\mathrm{ln}[1+e^{-\beta(\varepsilon-\mu)}]$. Therefore, the oribal magnetization as well as polarization can be derived as
\begin{equation}
 M (E)=-\frac{\partial F(E,B)}{\partial B}|_{B=0}, 
\end{equation} 
and
\begin{equation}
 P(B)=-\frac{\partial F(E,B)}{\partial E}|_{E=0}.
\end{equation} 
We can therefore obtain the $M(E)$ as well as $P(B)$ as
\begin{equation}
M(E)=\sum_{n\bm{k}}m_{n\bm{k}}f_{n\bm{k}}+\frac{1}{\beta}\frac{e}{\hbar}\sum_{n\bm{k}}\Omega_{n\bm{k}}\phi\big[\varepsilon_{n\bm{k}} (E)\big]
\end{equation}
and 
\begin{equation}
P(B)=-\sum_{n\bm{k}}p_{n\bm{k}}f_{n\bm{k}}+\frac{e}{\hbar}B [\frac{1}{\beta}\phi(\varepsilon_{n\bm{k}})\delta_\Delta \Omega_{n\bm{k}}-\Omega_{n\bm{k}}p_{n\bm{k}}]
\end{equation}
From the above derivation, we can see that the magnetic field can give an additional term to the electric polarization. We further do the linear expansion and obtain the LMC coefficients as
\begin{equation}
\label{eq:generalchi}
\begin{split}
\chi_{me} &=\chi_{em} =  e \sum_{n\bm{k}} \Big\{\delta_\Delta m_{n\bm{k}} f_{n\bm{k}} +  m_{n\bm{k}} p_n f'_{n\bm{k}}  +\\
&(e/\hbar) \big(\delta_\Delta \Omega_{n\bm{k}} \phi \big[\varepsilon_{n\bm{k}}\big]/\beta - \Omega_{n\bm{k}} f_{n\bm{k}} p_n\big) \Big\}
\end{split}
\end{equation}
From this derivation we can clearly see that the orbital magnetoelectricity is a bulk property containing both the orbial magnetic moment and the Berry phase correction. 

In the above derivation, we focus on the intraband contribution, while the interband correction is
\begin{equation}
\chi_{me}^{inter}=2\sum_{l\neq n,\bm{k}}\frac{Re(m_{nl}p_{ln})}{\varepsilon_{nl}}f_{n\bm{k}}
\end{equation}
Therefore, we can define the orbital ME moment $\alpha_{n\bm{k}}$ as
\begin{equation}
\alpha_{n\bm{k}}=\delta_\Delta m_{n\bm{k}}+2\sum_{l\neq n,\bm{k}}\frac{Re(m_{nl}p_{ln})}{\varepsilon_{nl}}
\end{equation}
The final result can be written as the summation of the intraband and interband contributions, which reads
\begin{equation}
\begin{split}
\chi_{me} &= e \sum_{n\bm{k}} \Big\{\alpha_{n\bm{k}}f_{n\bm{k}} +  m_{n\bm{k}} p_n f'_{n\bm{k}}  \\
&+ (e/\hbar) \big(\delta_\Delta \Omega_{n\bm{k}} \phi \big[\varepsilon_{n\bm{k}}\big]/\beta - \Omega_{n\bm{k}} f_{n\bm{k}} p_n\big) \Big\}
\end{split}
\end{equation}
which is consistent with Ref.~\cite{hu2026orbital}. In this work, we do not consider the interband correction since it's a high order term.

%

\end{document}